\documentclass[conference]{IEEEtran}
\usepackage{amsmath,amssymb,amsfonts}
\usepackage{algorithmic}
\usepackage{graphicx}
\usepackage{textcomp}
\usepackage{xcolor}
\def\BibTeX{{\rm B\kern-.05em{\sc i\kern-.025em b}\kern-.08em
    T\kern-.1667em\lower.7ex\hbox{E}\kern-.125emX}}

\usepackage{enumitem} 
\usepackage{siunitx}  
\usepackage{xspace}
\usepackage{caption}
\usepackage{float}
\usepackage{listings}
\usepackage[dvipsnames]{xcolor} 
\usepackage{url}                            
\usepackage{hyperref}                       
\usepackage[nameinlink]{cleveref}
\usepackage{booktabs}
\usepackage[most]{tcolorbox}
\usepackage{array}
\usepackage{pifont} 
\usepackage{multirow}
\usepackage[numbers,sort&compress]{natbib}
\usepackage{colortbl}
\usepackage{dsfont}    
\usepackage{amsthm}

\newcommand{\Fone}{F1}

\newtheorem{definition}{Definition}
\definecolor{mildred}{RGB}{220, 90, 90}  
\definecolor{mildblue}{RGB}{90, 130, 200} 

\hypersetup{
    colorlinks=true,
    linkcolor=blue,
    citecolor=blue,
    filecolor=magenta,
    urlcolor=blue
}

\begin{document}

\title{Detecting Semantic Clones of Unseen Functionality}

\author{
\IEEEauthorblockN{Konstantinos Kitsios}
\IEEEauthorblockA{
\textit{University of Zurich}\\
Zurich, Switzerland \\
konstantinos.kitsios@uzh.ch}
\and
\IEEEauthorblockN{Francesco Sovrano}
\IEEEauthorblockA{
\textit{University of Zurich}\\
Zurich, Switzerland \\
francesco.sovrano@uzh.ch}
\and
\IEEEauthorblockN{Earl T. Barr}
\IEEEauthorblockA{
\textit{University College London} \\
London, UK \\
e.barr@ucl.ac.uk}
\and
\IEEEauthorblockN{Alberto Bacchelli}
\IEEEauthorblockA{
\textit{University of Zurich}\\
Zurich, Switzerland \\
bacchelli@ifi.uzh.ch}
}

\maketitle

\begin{abstract}
Semantic code clone detection is the task of detecting whether two snippets of code implement the same functionality (e.g., Sort Array). Recently, many neural models achieved near-perfect performance on this task.
These models seek to make inferences based on their training data. 
Consequently, they better detect clones similar to those they
have seen during training and may struggle to detect those they have not.  
Developers seeking clones are, of course, interested in both types of clones.
We confirm this claim through a literature review, identifying three practical clone detection tasks in which the model's goal is to detect clones of a functionality even if it was trained on clones of different functionalities.
In light of this finding, we re-evaluate six state-of-the-art models, including both task-specific models and generative LLMs, on the task of detecting clones of unseen functionality.
Our experiments reveal a drop in \Fone{} of up to 48\% (average 31\%) for task-specific models. 
LLMs perform on par with task-specific models without explicit training for clone detection, but generalize better to unseen functionalities, where \Fone{} drops up to 5\% (average 3\%) instead. 

We propose and evaluate the use of contrastive learning to improve the performance of existing models on clones of unseen functionality. 
We draw inspiration from the computer vision and natural language processing fields where contrastive learning excels at measuring similarity between two objects, even if they come from classes unseen during training. 
We replace the final classifier of the task-specific models with a contrastive classifier, while for the generative LLMs we propose contrastive in-context learning, guiding the LLMs to focus on the differences between clones and non-clones. 
The \Fone{} on clones of unseen functionality is improved by up to 26\% (average 9\%) for task-specific models and up to 5\% (average 3\%) for LLMs.

\noindent \textit{Data and material: \href{https://doi.org/10.5281/zenodo.17238379}{https://doi.org/10.5281/zenodo.17238379}}
\end{abstract}


\begingroup\renewcommand\thefootnote{}\footnote{
This is the author’s version of the paper accepted for publication in the 40th IEEE/ACM International Conference on Automated Software Engineering (ASE 2025). The final version will be available via IEEE Xplore.}
\addtocounter{footnote}{-1}\endgroup

\section{Introduction}\label{sec:introduction}
Code clones are snippets of code that implement the same functionality;
this functionality is sufficiently intricate that a developer prefers to reuse the code snippet rather than regenerate it from scratch~\cite{functionality, functionality_non_trivial}. 
It is this latter constraint that separates clones from single statements or other short programming idioms.
Clones are typically introduced through copy-and-paste~\cite{roy2007survey}. 
Although clones can be used to reduce coupling and prepare subsequent development~\cite{kapser2008cloning}, they can also have negative consequences, including increased maintenance costs~\cite{roy2007survey,rattan2013software}, reduced software quality~\cite{cloneSmells}, increased risk of introducing bugs~\cite{roy2007survey,doClonesMatter}, reduced code reusability~\cite{tsantalis2015assessing}, and difficulty in modifying the code~\cite{roy2007survey}.

If two code snippets are syntactically similar they are called \textit{syntactic clones}, otherwise they are called \textit{semantic clones}~\cite{roy2007survey}. 
For example, a recursive and a dynamic-programming implementation of a function that returns the $n^{th}$ Fibonacci number are semantic clones. 
Syntactic clone detection is typically solved with
token-matching and tree-matching algorithms~\cite{svajlenko2015evaluating}. 
On the other hand, semantic clone detection is undecidable according to Rice's theorem~\cite{rice_theorem}, so we can only approximately solve it.
Following the rise of deep learning, numerous neural clone detection models have been proposed for this purpose~\cite{codebert,CDLH,astnn,codegrid,wysiwim,wang2020detecting,zeng2019fast,mlIsAllUNeed,xastnn}.
Neural clone detection models are trained in a supervised setting on task-specific datasets. 
The most widely used dataset is BigCloneBench (BCB)~\cite{bcb}, which contains \num{43} functionalities.
We refer to a \textit{functionality} as a non-obvious sequence of statements to accomplish a clearly and compactly definable program task~\cite{functionality,functionality_non_trivial}, e.g., \texttt{Sort Array} or \texttt{Fibonacci}.
For each one of the \num{43} functionalities, BCB contains both clone pairs (two code snippets that both implement the functionality) and non-clone pairs (two code snippets, only one of which implements the functionality).

For the training and evaluation of neural models, the functionalities in the dataset are split into training and test sets uniformly at \textit{random}, leading to some clone pairs of the \texttt{Fibonacci} functionality ending up in the training set and some others in the test set (see  \Cref{fig:evaluation_methods}a). 
In this setting, state-of-the-art (SOTA) models achieve outstanding results of up to $97.8\%$ \Fone~\cite{codegrid}.
This near-perfect \Fone{} concerns the task of detecting clones of functionalities whose clones also appear in the training set.
For example, the models excel at the task of detecting clones of the \texttt{Fibonacci} functionality when they are trained on clones of the \texttt{Fibonacci} functionality.
Although this task is meaningful, we investigate the importance of the task of detecting clones of the \texttt{Fibonacci} functionality when the model is trained on clones of functionalities \emph{other than} \texttt{Fibonacci}.
In the former case, we say that clones of the \texttt{Fibonacci} functionality are \textit{seen during training} while in the latter case, \textit{unseen during training}.
Through an initial motivation study, we find that detecting clones of unseen functionality is a more practical, naturally-occurring task, yet, SOTA models are evaluated on clones of seen functionality.

This motivates us to evaluate and improve the performance of SOTA models on this task.
To this aim, we evaluate the performance of three task-specific models and three generative LLMs on clones of unseen functionality, measuring an \Fone{} of up to 71.5\% for task-specific models and up to 69\% for LLMs. 
LLMs are more robust at detecting clones of unseen functionalities, with an average \Fone{} drop of 3\% compared to their \Fone{} on seen functionalities, while the same drop is 31\% for task-specific models.

The performance drop on clones of unseen functionality motivates us to devise techniques that increase the performance of existing models on this task.
We propose and evaluate the use of contrastive learning (CL), drawing inspiration from the computer vision and natural language processing fields where measuring similarity between two objects emerges in tasks like face verification and sentence similarity. 
Recent work in these fields shows the superiority of CL~\cite{siamese_first_paper} because its training explicitly optimizes for similarity~\cite{koch2015siamese,sentence_similarity,allamanis2020typilus,siamese6}. 
Moreover, Koch et al.~\cite{koschke2012licence} showed that CL generalizes well to classes \emph{unseen during training}.
Thus, we investigate whether contrastive learning can increase the performance of models on clones of unseen functionality. 
We replace the final classifier of the task-specific models (e.g., CodeBERT trained for clone detection) with a contrastive classifier. For generative LLMs, we propose an analogous prompting technique that guides the LLM toward the differences between clone and non-clone pairs.
Results show increased performance in \num{9} out of \num{12} experiments and no difference in \num{3} experiments.
We qualitatively analyze an experiment where CL did not increase performance and find that CL led to learning a different, more strict definition of a clone.
Our work led to the following main research contributions:
\begin{itemize}
    \item empirical evidence of a significant performance drop of six SOTA models on clones of unseen functionality;
    \item a dataset and an evaluation methodology for evaluating future clone detectors on clones of unseen functionality;
    \item empirical evidence that CL in post-training improves the performance of state-of-the-art task-specific models on clones of unseen functionality;
    \item contrastive clone prompting, a novel prompting technique that improves the performance of state-of-the-art generative LLMs on clones of unseen functionality.
\end{itemize}

\begin{table}
    \centering
    \scriptsize
    \caption{Summary of the datasets employed in our study.}\label{table:datasets}
    \begin{tabular}{lllll}
        \toprule
        Dataset & \# unique funct. & \# clone pairs & \# non-clone pairs & Language \\
        \midrule
        BCB$_s$ & 43 & 8.6M & 258K & Java \\
        BCB$_{s'}$ & 23 & 2300 & 2300 & Java \\
        SCB & - & 1K & 1K & C \& Java \\
        OJ & 104 & 3K & 47K & C \\
        \bottomrule
    \end{tabular}
\end{table}

\section{Background and Related Work}\label{sec:background}
We present here the background of the datasets and models used in our experiments, and we also discuss related work.
In \Cref{ssec:rq1}, we justify the selection of these models among other state-of-the-art.

\subsection{Background}
\noindent\textbf{Datasets.}\label{ssec:datasets}
BigCloneBench (BCB) is the most widely used benchmark for clone detection (see \Cref{ssec:motivation}) and is built on Java. 
The initial version BCB\textsubscript{v1}~\cite{bcb} starts with the authors selecting ten target functionalities frequently encountered in Java projects.
For each functionality, they define search heuristics\footnote{The list of heuristics is available in the \href{https://github.com/clonebench/BigCloneBench?tab=readme-ov-file}{PostgreSQL version of the dataset}.} and apply them to a large corpus of open-source Java code~\cite{ija_dataset} to retrieve code snippets that \textit{potentially} implement the functionality. 
The retrieved code snippets are manually inspected by humans to determine if they actually implement the intended functionality.
Furthermore, the authors write their own code snippets that implement the intended functionality.
Finally, all the pairs that implement the intended functionality are clone pairs while the rest of the pairs within the intended functionality are non-clone pairs.
This results in \num{6.2}M clone pairs and \num{258}K non-clone pairs.
BCB\textsubscript{v2}~\cite{bcbv2} contains \num{43} functionalities with \num{8.6}M clone pairs and \num{279}K non-clone pairs. 
The large number of pairs and the class imbalance led researchers~\cite{astnn,wysiwim,codegrid} to adopt a subset, which we call BCB$_{s}$, by sampling \num{20}K clone and \num{20}K non-clone pairs from BCB\textsubscript{v2}.

SemanticCloneBench (SCB)~\cite{scb} is built by mining Stack Overflow, considering two correct answers to the same question as a clone pair. 
It contains \num{4000} clone pairs uniformly distributed across four languages (Java, Python, C, C\#).\footnote{We consider the \num{1000} Java pairs only, since BCB is also built on Java.} The dataset does not contain non-clone pairs, so researchers~\cite{codebert} artificially created \num{1000} non-clone pairs by combining answers to different questions. 

The OpenJudge (OJ) dataset~\cite{ojclone} consists of correct solutions to $104$ C problems (functionalities). Two programs that implement the same functionality are a clone pair; non-clone pairs are created by pairing programs that implement different functionalities. The version used in the literature consists of $50$K pairs, $93$\% of which are non-clones~\cite{liu2021can,astnn,xastnn,wysiwim}.
A summary of the datasets is shown in~\Cref{table:datasets}.

\smallskip
\noindent\textbf{Models.}
CodeBERT~\cite{codebert_model} is a general-purpose model for code. For semantic clone detection~\cite{codebert,codexglue,graphcodebert}, pairs of snippets separated by a \texttt{[SEP]} token are given as input and the model outputs a vector representation for each pair. 
The representation is fed into a fully-connected layer with softmax activation~\cite{backpropagation} to predict if the snippets are clones or not. 

\begin{figure}
  \centering
  \includegraphics[width=0.5\textwidth]{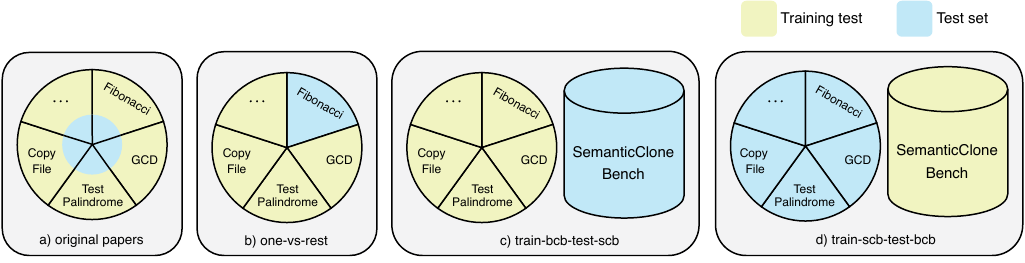}
  \caption{Overview of the train/test splits used in four evaluation methods. Pie slices represent functionalities of the BCB dataset and the cylinder represents the SCB dataset.
  Although split a) is used in the literature, only the splits b), c), and d) measure the cross-functionality performance.}
  \label{fig:evaluation_methods}
\end{figure}

ASTNN~\cite{astnn} is a code representation model designed to address the vanishing gradient issues caused by long ASTs. 
It achieves this by breaking down the AST of a given code snippet into smaller AST sequences, which are individually encoded and then fed into a bidirectional RNN that produces the representation.
Despite using RNNs---that are outperformed by Transformers~\cite{vaswani2017attention} in nearly every task~\cite{transformerVsRnn}---it still achieves near SOTA performance on software engineering tasks.
For semantic clone detection, the representations of the two snippets are subtracted and the result is fed into a fully-connected layer with sigmoid activation~\cite{backpropagation} to predict whether they are clones or not.

CodeGrid~\cite{codegrid} is a code representation model that leverages the spatial information of code. A code snippet 
is encoded using a token encoder (e.g. Word2Vec~\cite{mikolov2013word2vec}). 
The encoding of each token is placed in a grid that preserves the code layout, which is fed into a Convolutional Neural Network (CNN) to obtain the representation. 
For semantic clone detection, the representations of the two snippets are subtracted and the result is fed into a Support Vector Machine (SVM) classifier~\cite{svm} to predict whether they are clones or not. 

The three models above are \textit{task-specific} classification models that are trained on clone datasets to classify clones.
Recently, generative large language models (LLMs) have excelled in classification tasks without being explicitly trained on the classification objective~\cite{clones_llm_fewshot, llm_clones_crosslang}. They achieve this through in-context learning where the models are prompted with a few examples of clones before presenting the pair under classification. We study three such models, shown in \Cref{tab:llm_results}.

\subsection{Related Work}\label{sec:relatedWork}

We present here prior work related to contrastive learning for clone detection and generalizable clone detection.

\smallskip
\noindent\textbf{Contrastive Learning for Clone Detection.}
Contrastive learning has been applied in prior clone detection research: 
ContraCode~\cite{siamese1} is a pretrained code model trained with CL for clone detection.
Sia-RAE~\cite{siarae} detects clones by linking two recursive autoencoders with a comparator network while Zubkov et al.~\cite{siamese3} compare different paradigms of CL for clone detection. 
SJBCD~\cite{siamese4} is a clone detector trained using CL on Java bytecode instead of raw code. 
C4~\cite{siamese8} combines CodeBERT with CL to detect clones across different programming languages, while CC2Vec applies CL on two self-attention layers to efficiently generate token representations.
Contrastive learning has also been used in the pretraining stage, e.g., in UniXcoder~\cite{guo-etal-2022-unixcoder} and ContraBERT~\cite{contrabert}, where it outperformed the baseline training method on clone detection.

The goal of these works is to develop effective CL architectures.
As such, they are complementary to our goal, which
is to increase the performance of existing models in the unseen functionality setting that naturally appears in real-world tasks.
Our findings highlight the importance of these works even more, making CL a prominent go-to architecture for future clone detectors.
Additionally, our work goes further than task-specific models by proposing a contrastive prompting technique for in-context learning with LLMs. 

\smallskip
\noindent\textbf{Generalizable Clone Detection.}
Three studies have investigated cross-functionality clone detection.
Arshad et al.~\cite{codebert} aim to measure the cross-functionality performance of CodeBERT by training on all the functionalities of BCB and testing on SCB, reporting a drop in \Fone{} of $41\%$. 
Liu et al.~\cite{liu2021can} partition the functionalities of OJ in 7 parts, train two AST-based models~\cite{astnn,CDLH} in 1 part and test it in the other 6, reporting weak cross-functionality performance. 
Khajezade et al.~\cite{cross_problem_language} also follow the 7-part setting and augment it with cross-language evaluation.
Our contribution differs from these previous works in three ways. 
First, we present a novel systematic reasoning about the importance of cross-functionality performance in real-world applications, as we found no such reasoning in the literature.
Second, we evaluate both task-specific models and generative LLMs, highlighting the differences in their generalization abilities.
Third, we propose techniques to partially mitigate the observed performance drop, which only Khajezade et al.~\cite{cross_problem_language} do for task-specific models, while we also extend to generative LLMs.

Task-specific clone detectors have also been evaluated in cross-project~\cite{crossproject_clones}, cross-dataset~\cite{crossdataset,crossdataset2}, and cross-language~\cite{crosslanguage,crosslanguage2,llm_clones_crosslang} settings, which are complementary to cross-functionality: 
two different projects may share functionality, while many distinct functionalities exist within a single project, and the same holds for cross-language and cross-dataset.

\section{Motivation: Clones of Unseen Functionality are Overlooked}
\label{ssec:motivation}

Our work aims to evaluate and narrow the gap between semantic clone detection research and its practical applications. To this end, we study the occurrence of unseen functionalities---introduced informally in~\Cref{sec:introduction} and defined formally in~\Cref{sec:methodology}---in (a) real-world tasks, and (b) the evaluation of SOTA models proposed in recent literature.
We find that clones of unseen functionality appear in real-world tasks and yet have been overlooked in the literature.

\smallskip
\noindent\textbf{First-Principles Argument.} To investigate the relevance of detecting clones of unseen functionality in real-world applications, we begin with a first-principles combinatorial argument.
We claim that when generating source code in the order of millions of lines of code
(LOC), the creation of new, unseen functionalities is unavoidable.
Researchers have found that clones make up a big proportion of software systems, ranging from 7\% to 23\%~\cite{baxter1998clone,clone_perc,clone_perc2,clone_perc3}.
This large number of clones can be explained by only one of the following propositions:
\begin{enumerate}[label=(P\arabic*)]
    \item many clones of \emph{few} functionalities exist, or
    \item many clones of \emph{many} functionalities exist.
\end{enumerate}

We claim that (P1) cannot hold:
In the study of the uniqueness of source code~\cite{gabel2010study}, Gabel and Su show that (i) blocks of code with length greater than seven LOC are mostly ``unique''.
From \Cref{def:functionality} and \Cref{def:clones}, two blocks of code are clones only if the functionality they implement is intricate enough that it makes sense for the developer to reuse it rather than rewrite it from scratch.
Hence, (ii) we assume that most cloned functionalities in real-world projects are around or beyond the seven LOC threshold.
Combining (i) and (ii), we conclude that most of the cloned functionalities in real-world projects are unique functionalities.
Hence, we claim that (P2) holds, and many cloned functionalities exist in software systems.
Given that the number of functionalities in the training set of a clone detector ranges from \num{10} (BCB\textsubscript{v1}) to \num{104} (OJ), we conclude that \emph{functionalities unseen during training will dominate real-world software systems}, and deployed models will be required to detect clones of unseen functionalities.

\smallskip
\noindent\textbf{Literature Review of Real-World Applications.} We buttress our first-principles argument with evidence from our small-scale literature review on real-world applications of clone detection. 
We first search for papers whose title contains the keyword \texttt{clone} in the last ten years of ICSE, both in the main and Software Engineering in Practice tracks. 
Note that our goal is not an extensive literature review, but rather to provide evidence that, in real-world applications, a model should detect clones not only of seen functionalities, but also of new, unseen functionalities. 
We find \num{15} papers, of which we keep those that present an application of clone detection, discarding those that only evaluate a tool against a benchmark.
This results in only one paper, so we apply the same criteria to the papers mentioned in two well-established surveys by Roy et al.~\cite{roy2007survey} and Svajlenko et al.~\cite{svajlenko2020survey}. This leads to a total of three papers that satisfy our criteria, which we analyze here.

Ishihara et al.~\cite{ishihara2012libraries} run syntactic clone detection to a collection of \num{13}K open-source Java projects.
Their goal is to locate frequently cloned functionalities to implement them as external libraries. 
They found \num{56} such functionalities, an example being the sorting of a JTable, which was independently added to the Java SE \num{6}. 
To succeed at this task, a clone detector should detect clones of table sorting algorithms even if it has not seen such clones in the training set. 

In another application, Laguë et al.~\cite{lague1997assessing} studied communications software and found empirical evidence regarding the benefits of \textit{Preventive Control}, which involves the use of a clone detector in the development process to prevent new clones from entering the system. 
In the modern code review process~\cite{bacchelli2013expectations}, this is implemented by integrating a clone detector in the pipeline before merging a pull request and many commercial tools support it (e.g., \href{https://www.sonarsource.com/blog/manage-duplicated-code-with-sonar/}{Sonar}). 
If the clone detector only detects clones of functionalities it has seen in the training, the reported performance of up to \num{98}\% \Fone{} is an overestimation.

Microsoft used syntactic clone detection to find security vulnerabilities~\cite{xiao,xiao2}.
Once developers discovered a vulnerability, they searched for clones of the vulnerable code.
For example, developers discovered a snippet that could cause potential heap corruption. After searching for clones of this snippet, they found three more snippets that could cause similar heap corruptions.
The clone detector should not only find clones of known vulnerable snippets but also unseen ones to be useful in this scenario.
Although the study focused on syntactic clones, the authors stated their intention to pursue semantic clones in future work~\cite{xiao2}.
The evidence from our literature review provides substantial support to our first-principles reasoning, and by combining them we reach the following observation:

\begin{tcolorbox}[colback=gray!8!white, colframe=gray!60!black, boxrule=0.5pt, arc=2pt, left=6pt, right=6pt, top=4pt, bottom=4pt]
\begin{minipage}{\textwidth}
\textbf{Observation 1.} We find three applications where detecting clones of unseen functionality is the core task.
\end{minipage}
\end{tcolorbox}

\smallskip
\noindent\textbf{Evaluation Methods Used in the Literature.} Then, we investigate the evaluation methods applied in the literature.
We search in the last five years of the ICSE for neural models that achieve SOTA performance on clone detection, and also include models from top software engineering venues that reference them as a baseline and outperform them.
We find nine models whose details are shown in \Cref{table:models}. 
We group the models by their input format: text-based models extract representations from raw code, while AST-based models feed the Abstract Syntax Tree (AST) of the code into RNNs or Graph Neural Networks for this purpose. 
Image-based approaches extract representations by feeding the image signal of code into CNNs.
Finally, GraphCodeBERT~\cite{graphcodebert} accepts hybrid input, i.e., it combines text information with data flow.

By analyzing the evaluation methods used in the nine models of~\Cref{table:models}, we extract the following information.
The most frequently used dataset is BCB (9 papers), followed by OJ (4 papers) and GoogleCodeJam (GCJ)~\cite{faast} (1 paper). 
All three datasets are functionality-based, i.e., they consist of a number of functionalities (43, 104, 12, respectively) and contain clones of each functionality.
\emph{All the models are evaluated on the task of detecting clones of seen functionalities}: 
the datasets are split uniformly at random into training and test sets, resulting in functionalities of the test set appearing in the training set, as shown in ~\Cref{fig:evaluation_methods}a.

\begin{tcolorbox}[colback=gray!8!white, colframe=gray!60!black, boxrule=0.5pt, arc=2pt, left=6pt, right=6pt, top=4pt, bottom=4pt]
\begin{minipage}{\textwidth}
\textbf{Observation 2.} The nine studied state-of-the-art models are evaluated on the task of detecting clones of \textit{seen} functionality.
\end{minipage}
\end{tcolorbox}

Our systematic search of the last five years of ICSE did not yield any generative LLMs, however, such models are effective in classification tasks through in-context learning~\cite{icl}, even surpassing task-specific models at times~\cite{llm_clones_crosslang}.
Since the relevance of LLMs for software engineering tasks is rising, we take up the task of studying the effect of unseen functionalities in the in-context examples of LLMs as well, laying the ground for future evaluations of LLMs for clone detection.

\section{Methodology}\label{sec:methodology}
First, we formally define the main concepts of this paper. 

\begin{definition}\label{def:functionality}
A \textbf{functionality} is a non-obvious sequence of code
statements to accomplish a clearly and compactly definable task.
\end{definition}

Examples of functionalities are \texttt{Fibonacci} (calculate the $n^{th}$ Fibonacci number), \texttt{GCD} (Calculate the Greatest Common Divisor of two integers), or \texttt{Test Palindrome} (check if a string reads the same from both ends).
The term \textit{``non-obvious''} differentiates a functionality from single statements or short programming idioms.
This makes a functionality intricate enough that a developer prefers to reuse the code snippet rather than regenerate it from scratch.

\begin{definition}\label{def:clones}
Two code snippets form a \textbf{clone pair} iff they implement the same functionality.
\end{definition}

A member of a clone pair is called a clone.
By negation, two code snippets that do not implement the same functionality form a non-clone pair.
While intuitive, this definition of a clone is undecidable because it rests on code equivalence~\cite{rice_theorem}.
It also leaves intentionally imprecise whether two snippets are clones when they implement unshared---in addition to shared---behavior, an issue we take up in \Cref{ssec:robustness_testing}.
It does, however, align with the definition used in the literature~\cite{bcb}.

\begin{definition}\label{def:unseen_functionality}
A clone in the test set is a clone of an \textbf{unseen functionality}, if the model has not explicitly learned to detect clones of this functionality.
\end{definition}

The term \textit{learning} in the above definition includes both training for \textit{task-specific models} and in-context learning for \textit{LLMs}.
For task-specific models, if the training set contains pairs of the functionalities \texttt{Test Palindrome} and \texttt{GCD} and the test set contains pairs of \texttt{Fibonacci}, we say that the latter are clones of an unseen functionality.
This definition considers as unseen a functionality also in the case where a snippet that implements it appeared in the pretraining corpus of LLMs (e.g., GPT-4o). However, the LLMs must not have  learned \textit{explicitly} to classify clone pairs of the said functionality, e.g., through finetuning or in-context learning.

We refer to the task of detecting clones of unseen functionality as \textit{cross-functionality} clone detection and to the performance of models in this task as cross-functionality performance.

\subsection{Research Questions}

The study that motivates this work,  in~\Cref{ssec:motivation}, reveals that (a) detecting clones of unseen functionality occurs in applications and (b) SOTA models are evaluated on clones of \textit{seen} functionality.
This motivates the evaluation of models' cross-functionality performance.
To this aim, we evaluate three task-specific models and three generative LLMs on settings where the clone pair under classification comes from a functionality unseen in the finetuning dataset (for task-specific models) or the in-context examples (for generative LLMs).
\begin{tcolorbox}[colback=gray!8!white, colframe=gray!60!black, boxrule=0.5pt, arc=2pt, left=6pt, right=6pt, top=4pt, bottom=4pt]
\begin{minipage}{\textwidth}
\textbf{RQ1}. How well do state-of-the-art models detect clones of unseen functionality?
\end{minipage}
\end{tcolorbox}

To conclude, we propose and evaluate the use of contrastive learning to improve cross-functionality performance. 
We draw inspiration from adjacent research areas where CL showcases impressive performance in measuring similarity between two objects, even when they come from classes unseen during training. 
We adapt three task-specific models by replacing their final classifier
with a contrastive classifier and also propose an analogous prompting technique for LLMs, called contrastive in-context learning, to answer our last research question.
\begin{tcolorbox}[colback=gray!8!white, colframe=gray!60!black, boxrule=0.5pt, arc=2pt, left=6pt, right=6pt, top=4pt, bottom=4pt]
\begin{minipage}{\textwidth}
\textbf{RQ2}. To what extent does contrastive learning improve model performance on clones of unseen functionality?
\end{minipage}
\end{tcolorbox}

\begin{table}
    \centering
    \scriptsize
    \caption{Task-specific model's performance on clones of seen functionality. The models we evaluate in this work appear highlighted.}\label{table:models}
    \begin{tabular}{@{}lccccc}
        \toprule
        Model & Input format & \Fone{} & P & R & A \\
        \midrule
        \rowcolor{gray!20}
        CodeBERT~\cite{codebert} & Text   & 95.0 & 99.0 & 91.0 & 95.0 \\
        Toma~\cite{mlIsAllUNeed} & Text   & 90.0 & 93.0 & 88.0 & - \\
        \rowcolor{gray!20}
        ASTNN~\cite{astnn}       & AST    & 93.8 & 99.8 & 88.3 & -    \\
        xASTNN~\cite{xastnn}     & AST    & 96.6 & \textbf{99.9} & 93.5 & -    \\
        FA-AST~\cite{faast}      & AST    & 95.0 & 96.0 & 94.0 & -    \\
        CDLH~\cite{CDLH}         & AST    & 82.0 & 92.0 & 74.0 & -    \\ 
        WySiWiM~\cite{wysiwim}   & Image  & 94.8 & 95.3 & 94.3 & -    \\
        \rowcolor{gray!20}
        CodeGrid~\cite{codegrid} & Image  & \textbf{97.8} & 99.6 & \textbf{96.1} & -    \\
        GraphCodeBERT~\cite{graphcodebert} & Hybrid & 95.0 & 94.8 & 93.4 & - \\
        \bottomrule
    \end{tabular}
\end{table}

\subsection{RQ1 --- Cross-functionality Performance}\label{ssec:rq1}
The results of~\Cref{ssec:motivation} motivate us to quantify the cross-functionality performance of SOTA models.
To do so, we propose four evaluation methods designed to evaluate the models on clones of unseen functionality.
Model performance is measured by accuracy, \Fone{} score, precision, and recall.

\smallskip
\noindent\textbf{Model Selection.}
Experimenting with all nine models of \Cref{table:models} would be computationally intensive, so we group them by input format and select one model from each group.
CodeBERT is the best-performing \emph{text-based model} so we select it. 
From the four \emph{AST-based models}, CDLH and FA-AST are evaluated on a version of BCB that was later marked as problematic by Krinke et al.~\cite{bcb_harmful}, who found that the version contains incorrectly labeled non-clone pairs.
From the remaining two models, xASTNN does not come with a replication package, so we select ASTNN.
From the \emph{image-based models}, we select the best performer (CodeGrid).
Finally, GraphCodeBERT was also evaluated on the same problematic version of BCB~\cite{bcb_harmful}, so we do not use it in our experiments.
We end up with three models, which are highlighted in \Cref{table:models}.
Regarding generative LLMs, we select three SOTA models shown in \Cref{tab:llm_results}, covering commercial, open-source, and reasoning models.

\begin{figure}
  \centering
  \includegraphics[width=0.45\textwidth]{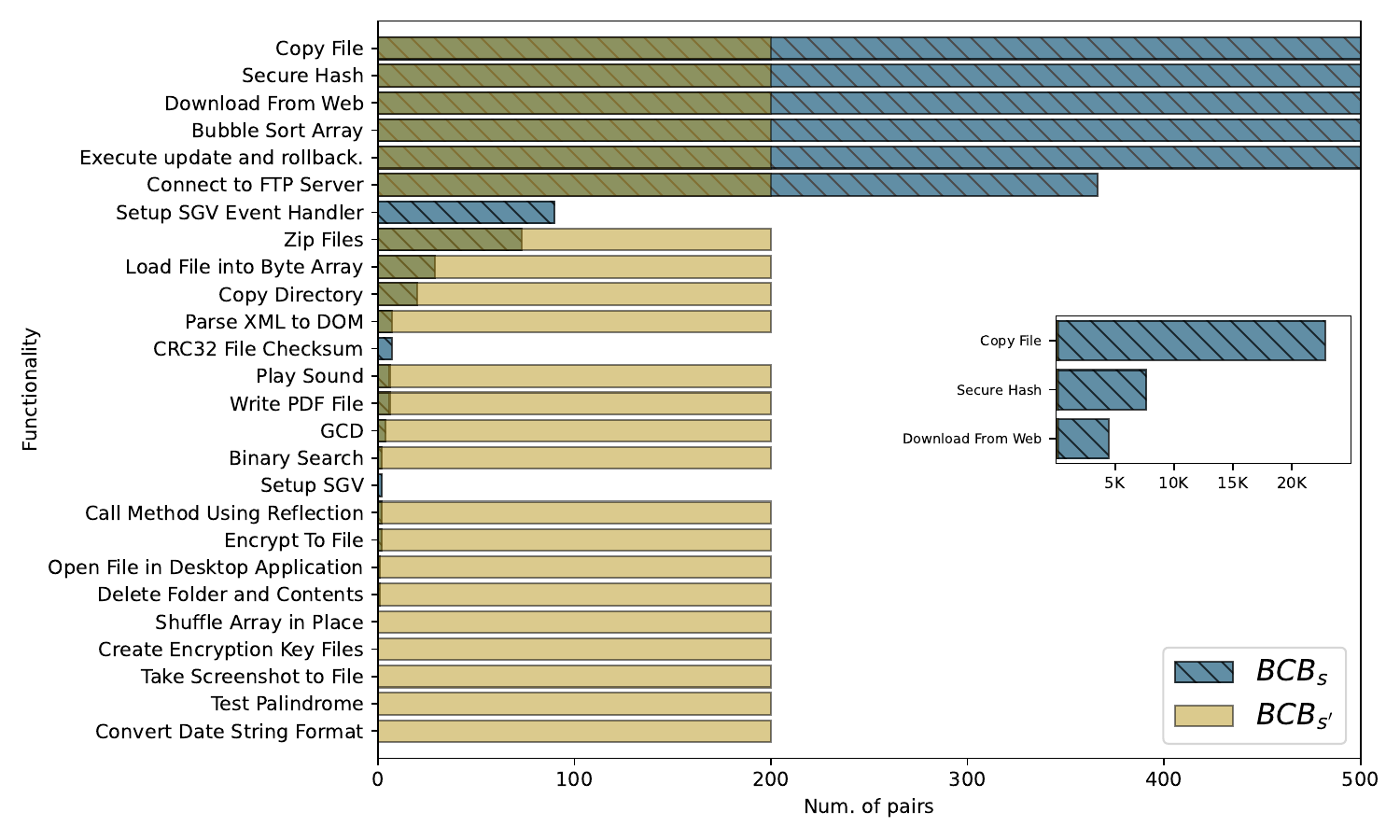}
  \caption{Number of pairs in each functionality of the two BCB dataset versions. The top-3 functionalities (also shown in the inset axis) make up $90\%$ of BCB$_{s}$.
  }

  \label{fig:bcb_distribution}
\end{figure}

\smallskip
\noindent\textbf{Dataset Selection.}
The models are evaluated on the BCB dataset in their original papers, so the first dataset we use is BCB. 
To increase the generality of our results, we also use the OJ dataset (\Cref{ssec:robustness_testing}).
In the literature, different deviations from the original BCB are used.
For example, three papers evaluate CodeBERT for code clone detection, with two of them~\cite{codexglue,graphcodebert} using the problematic version of BCB~\cite{bcb_harmful}.
Hence, for this work, we adopt the results from the third paper~\cite{codebert} that uses a custom subset of BCB\textsubscript{v2} of size \num{1000}. 

ASTNN and CodeGrid are evaluated on BCB$_{s}$ which was sampled from BCB\textsubscript{v2}. BCB\textsubscript{v2} is imbalanced in terms of functionalities: Although it contains \num{43} functionalities, $70\%$ of the non-clone pairs and $54\%$ of the clone pairs belong to the \texttt{Copy File} functionality alone~\cite{bcb_harmful}.
By analyzing BCB$_{s}$ we come to similar findings. 
The \texttt{Copy File} functionality makes up $60\%$ of the dataset and the three most frequent functionalities 
make up $90\%$ of the dataset. 
Moreover, only \num{10} out of the \num{43} functionalities have at least ten pairs in BCB$_{s}$. 
The full distribution is shown in blue in~\Cref{fig:bcb_distribution}.

To mitigate the functionality imbalance, we introduce \linebreak BCB$_{s'}$ that is obtained from BCB\textsubscript{v2} by performing functionality-aware sampling and thresholding: 
If a functionality has more than $N$ pairs, we sample $N$ of them uniformly at random so that functionalities like \texttt{Copy File} do not overshadow the others. 
If a functionality has fewer than $N$ pairs, we discard it. 
Finally, we ensure class balance by selecting $M=\frac{N}{2}$ clone pairs and $M$ non-clone pairs, similarly to BCB$_{s}$.
By selecting $M=100$, the thresholding retains $23$ functionalities, while for larger $M$ the number of functionalities decreases steeply.
Thus, the dataset contains $100$ clone pairs and $100$ non-clone pairs for each of the $23$ functionalities, for a total of \num{4600} pairs.
The new distribution of the retained functionalities is shown in~\Cref{fig:bcb_distribution}.
We use a random seed for the sampling and publicly release the code for creating the dataset to ensure reproducibility. 
BCB$_{s'}$ mitigates the imbalance threat that comes with BCB$_{s}$~\cite{bcb_harmful} so we use this version of BCB in our experiments, and also encourage its use in future research.

\smallskip
\noindent\textbf{Evaluating Task-Specific Models.}
We propose the three evaluation methods of ~\Cref{fig:evaluation_methods} to capture cross-functionality performance of task-specific models that require explicit training.
We start with the \textit{one-vs-rest} evaluation. 
Given a dataset with $F$ functionalities denoted by $\mathcal{F} = \left\{1, 2, ..., F\right\}$, we propose a sequence of $F$ experiments where, in experiment $i$, the training set consists of code pairs that implement the functionalities $\mathcal{F}-\left\{i\right\}$ and the test set consists of code pairs that implement the functionality $i$. 
BCB$_{s'}$ has $F=23$ functionalities, so we obtain $23$ tuples of (accuracy, \Fone{}, precision, recall) for each model. 
To compare the cross-functionality performance with the performance on clones of seen functionality, we employ the \textit{one-sample Wilcoxon signed-rank test}~\cite{wilcoxon} 
with the alternative hypothesis that the \num{23} \Fone{} values measured by the \textit{one-vs-rest} evaluation are lower than the single value reported in the original papers.

The \emph{one-vs-rest} evaluation satisfies the condition that functionalities in the test set do not appear in the training set; however, all functionalities come from the BCB$_{s'}$ dataset. 
To account for potential within-dataset biases, we also employ the \textit{train-bcb-test-scb} evaluation that uses all the functionalities of BCB$_{s'}$ for training and tests the trained model on the SCB dataset. 
We follow the exact permutation of Arshad et al.~\cite{codebert} to create non-clone pairs for SCB by combining code snippets that belong to different clone pairs, because they manually verified a fraction of the non-clone pairs. 
The resulting dataset is balanced and contains \num{2000} pairs.
Finally, we employ the complementary method \textit{train-scb-test-bcb} where the model is trained on SCB and tested on BCB$_{s'}$.

The hyperparameters of the models are kept unchanged for comparison with the original papers. 
Moreover, we only train the last layer of CodeBERT to avoid overfitting given the relatively small dataset size (e.g., \num{2}K for SCB).
We call the models CodeBERT\textsubscript{\textit{B}}, ASTNN\textsubscript{\textit{B}}, and CodeGrid\textsubscript{\textit{B}} respectively, where \textit{B} stands for Baseline, to distinguish them from the contrastive learning variations (RQ2) and report the results in \Cref{sec:results}.
Experiments were conducted on a Linux machine with an NVIDIA Tesla T4 GPU (16 GB memory) and an AMD CPU (4 cores, 32 GB memory).

\smallskip
\noindent\textbf{Evaluating Generative LLMs.}
The ability of generative LLMs to perform on par with task-specific models in classification comes with substantially higher cost.
To keep the cost of our LLM experiments viable, we run an a priori power analysis ($\alpha = 0.05, power = 0.8$) using Fisher's exact test to detect a small effect size. The analysis determined that 404 predictions are required to detect a change between two LLM prompts, i.e., baseline and contrastive.
This dataset size is typical in adjacent software engineering tasks~\cite{llms_vuln_detection,swtbench} that evaluate LLMs.
We sample \num{404} pairs of the BCB$_{s'}$ dataset uniformly at random and use them to evaluate the LLMs. 

LLMs are used in classification tasks like clone detection through in-context learning~\cite{clones_llm_fewshot}: they are prompted with clone and non-clone examples along with the new pair to classify.
To evaluate LLMs on clones of \textit{seen} functionality, we select the clone examples to have the same functionality as the pair under classification, while to evaluate them on \textit{unseen} functionality we select the clone examples from different functionalities.
We note that according to \Cref{def:unseen_functionality}, a functionality is unseen even if snippets implementing it were present in the pretraining stage of the LLMs; however, the LLM must not have explicitly learned to classify clones of these functionalities, e.g., through in-context learning.
We provide one clone and one non-clone pair to keep the prompt length small and use a temperature of 0 to enable  reproducibility,
following previous work~\cite{sovrano2025lost_in_the_end}.
We refer to this prompt as \textit{baseline}, because the clone and non-clone examples are selected uniformly at random, unlike the contrastive prompt in \Cref{ssec:siamese_architecture}.
An outline of the prompt is shown in \Cref{fig:prompt} (excluding the blue line), and we provide the full prompt in our replication package~\cite{ours}.
Through a system command, we required the LLM to generate the explanation prior to the decision, enabling the use of the explanation as a means of self-correction~\cite{self_correction}. Since our focus is on unseen functionalities and not on devising the optimal prompt, we do not experiment with alternative prompting schemes.
We present the results in \Cref{sec:results}.

\begin{figure}
  \centering
  \includegraphics[width=0.45\textwidth]{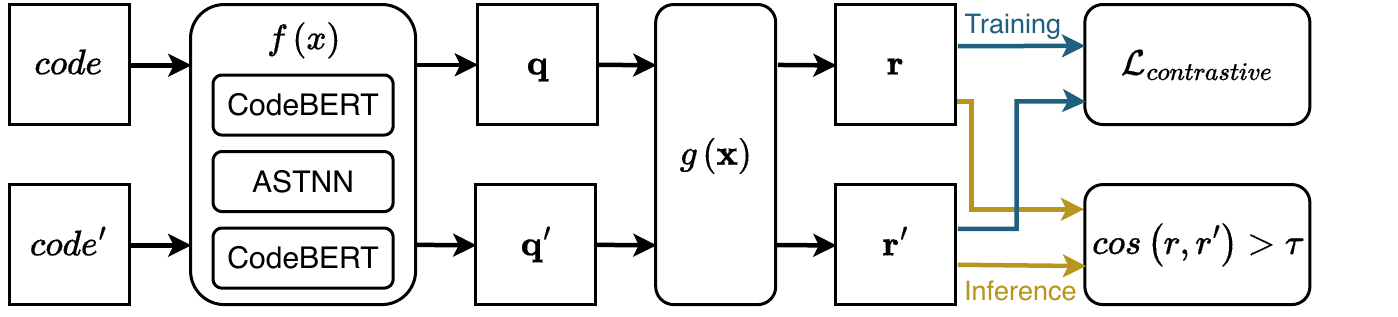}
  \caption{The contrastive learning architecture we use for code clone detection.}
  \label{fig:siamese_architecture}
\end{figure}

\begin{figure}
\centering
\scriptsize
\begin{tcolorbox}[colback=gray!10, colframe=black, width=0.49\textwidth, boxrule=0.5pt, left=6pt, right=6pt, top=4pt, bottom=4pt]
\textcolor{gray}{\texttt{Determine whether \{c\textsubscript{1}\},\{c\textsubscript{2}\} is a clone pair or not, based on the following definition: \{Definition\} \\
Clone example: \{\textbf{x}\},\{y\} \\
\textcolor{mildred}{Non-clone example: \{z\},\{w\} \# only in baseline} \\
\textcolor{mildblue}{ Non-clone example: \{\textbf{x}\},\{w\} \# only in contrastive }
}
}
\end{tcolorbox}
\caption{Prompt outline. In contrastive clone prompting, the clone and the non-clone example share the snippet \textbf{x}.}
\label{fig:prompt}
\end{figure}

\subsection{RQ2 --- Contrastive Learning Improvements}\label{ssec:siamese_architecture}
Contrastive learning~\cite{siamese_first_paper} is a neural network architecture designed for similarity learning that excels at identifying similar objects while discriminating between dissimilar ones. It explicitly maximizes the similarity of the vector representations of semantic clone pairs, in contrast to the baseline architectures described in \Cref{sec:background} that implicitly maximize this similarity, because their main objective is to minimize the classification loss function, hence being more prone to overfitting patterns seen during training~\cite{keshtmand2022understanding}.
CL is applicable to task-specific models that are trained on task-specific classification datasets; however, an analogous prompting technique called \textit{contrastive prompting} or \textit{in-context contrastive learning} has emerged for in-context learning with LLMs~\cite{llms_vuln_detection}. We discuss both techniques in more detail below. 

\smallskip
\noindent\textbf{Contrastive Learning.}
Contrastive learning training forces the representations of similar objects to converge in the representation space while driving the representations of dissimilar objects apart. 
We adopt the CL architecture shown in \Cref{fig:siamese_architecture} because it is straightforward and it was successfully used for clone detection before~\cite{siamese3}.
We do not experiment with other architectures since our aim is to demonstrate the effectiveness of CL in detecting clones of unseen functionality, and we and hope future work will elaborate on this.

The function $f\!\left(x\right)$ embeds the two snippets, $code$ and $code'$, in a representation space. 
Both $f\!\left(x\right)$, $code$, and $code'$ are model-agnostic, hence CL can be applied regardless of the model or the input format (text, AST, image). Consequently, if $f\!\left(x\right)$ captures semantic similarity, then CL captures semantic similarity too.
The intermediate representations $\mathbf{q}$ and $\mathbf{q}'$ are then propagated through  $g(\mathbf{x})$, which is a hyperparameter. 
The resulting representations $\mathbf{r}$ and $\mathbf{r}'$ are fed into the contrastive loss function that is minimized in the training phase:

{\small
\begin{equation} \label{eq:contrastive_loss}
\begin{aligned}
\mathcal{L} = & \frac{1}{2N} \sum_{i=1}^{N} y_i \left( \lVert \mathbf{r}_i - \mathbf{r}_i' \rVert \right)^2 + (1 - y_i) \left[ \max(0, m - \lVert \mathbf{r}_i - \mathbf{r}_i' \rVert) \right]^2
\end{aligned}
\end{equation}
}

In \Cref{eq:contrastive_loss}, 
\begin{itemize}
    \item the subscript $i$ denotes the $i^{th}$ pair of the dataset,
    \item $y_i$ is $1$ if $code_i$ and $code_i'$ are clones and $0$ otherwise,
    \item $N$ is the number of pairs in the dataset,
    \item $\lVert \cdot \rVert$ denotes the Euclidean norm, and
    \item $m$ is the margin that controls the minimum distance between dissimilar pairs and is a hyperparameter.
\end{itemize}

From \Cref{eq:contrastive_loss} we see that in the training phase, the representations of clone pairs ($y_i=1$) are attracted while those of non-clone pairs ($y_i=0$) are repelled. In the inference phase, the prediction for a new pair is given by \Cref{eq:prediction}:

{\small
\begin{equation} \label{eq:prediction}
\hat{y_i} = \begin{cases}
    0, & \text{if } \text{cos}(\mathbf{r}_i, \mathbf{r}_i') < \tau \\
    1, & \text{if } \text{cos}(\mathbf{r}_i, \mathbf{r}_i') \geq \tau
\end{cases}
\end{equation}
}
where $\cos(\mathbf{a}, \mathbf{b})$ denotes the cosine similarity and the threshold $\tau$ is set to $\tau = 0.5$ following previous work~\cite{astnn,codebert,codegrid,wysiwim}.

We replace the final classifier of the three task-specific models
with the contrastive classifier of \Cref{fig:siamese_architecture} and re-run the experiments of \Cref{ssec:rq1}. 
We call the resulting models CodeBERT\textsubscript{\textit{CL}}, ASTNN\textsubscript{\textit{CL}} and CodeGrid\textsubscript{\textit{CL}} respectively. We determine the hyperparameters $m$ and $g(\mathbf{x})$ using grid search while the rest of the hyperparameters remain the same as the baseline models. 
The grid search consists of three values for $m \in \{0.5, 5, 50\}$, covering the range of small, moderate, and big values, and two values for $g$: $g(\mathbf{x})= \mathds{1}$ (identity function) and $g(\mathbf{x})=BN$ (Batch Normalization), with the best combination being $(m=0.5, \,g(\mathbf{x})=BN)$.

\smallskip
\noindent\textbf{Contrastive In-Context Learning for LLMs.}
\Cref{eq:contrastive_loss} cannot be used with generative LLMs that are trained to predict the next token in a sequence. 
Instead, we use \textit{contrastive in-context learning}, a technique used in related tasks like vulnerability detection~\cite{llms_vuln_detection} where the LLM is provided with both a vulnerable code snippet and its fixed version before asking it to classify a new code as vulnerable or not. This way, the LLM focuses on the differences between vulnerable and non-vulnerable code, similar to contrastive learning.
Steenhoek et al.~\cite{llms_vuln_detection} find that contrastive in-context learning performs better than all other prompting schemes in vulnerability detection.
We adapt it for clone detection by providing one clone pair and one non-clone pair that share a common snippet, guiding the LLM to focus on the differences. Our baseline and contrastive prompts are summarized in \Cref{fig:prompt} and the full prompts are available in our replication package~\cite{ours}.

\section{Results}\label{sec:results}

In this section, we answer our research questions.
\begin{table}
    \centering
    \small
    \caption{Task-specific model performance across different evaluation methods (columns) on the BCB and SCB datasets. Models perform significantly worse on unseen functionalities (last three columns), and contrastive learning (CL) partially mitigates this drop.}\label{table:results}
    \resizebox{\columnwidth}{!}{%

    \begin{tabular}{@{\hskip 0.07in}l@{\hskip 0.07in}c@{\hskip 0.07in}c@{\hskip 0.07in}c@{\hskip 0.07in}c@{\hskip 0.07in}||c@{\hskip 0.08in}c@{\hskip 0.08in}c@{\hskip 0.08in}c@{\hskip 0.08in}|c@{\hskip 0.08in}c@{\hskip 0.08in}c@{\hskip 0.08in}c|c@{\hskip 0.08in}c@{\hskip 0.08in}c@{\hskip 0.08in}c}
        \toprule
         Evaluation $\rightarrow$& \multicolumn{4}{c}{on seen functionalities} &
        \multicolumn{4}{c}{one-vs-rest} & \multicolumn{4}{c}{train-bcb-test-scb} & \multicolumn{4}{c}{train-scb-test-bcb} \\
        \cmidrule(lr){2-5} \cmidrule(lr){6-9} \cmidrule(lr){10-13} \cmidrule(lr){14-17}
        Model $\downarrow$ & A & \Fone{} & P & R & A & \Fone{} & P & R & A & \Fone{} & P & R & A & \Fone{} & P & R \\
        \midrule
        CodeBERT\textsubscript{\textit{B}}~\cite{codebert} & 95.0 & 95.0 & 99.0 & 91.0 & 82.3 & 85.5 & \textbf{90.6} & 84.3 & 49.8 & 65.7 & \textbf{96.1} & 49.9 & 50.9 & 63.2 & \textbf{84.4} & 50.6 \\
        \rowcolor{gray!20}
        CodeBERT\textsubscript{\textit{CL}} &  &  &  &  & \textbf{82.6} & \textbf{85.7} & 89.2 & \textbf{86.3} & \textbf{84.8} & \textbf{84.1} & 89.0 & \textbf{81.0} & \textbf{67.1} & \textbf{67.6} & 68.8 & \textbf{66.5} \\
        \midrule
        ASTNN\textsubscript{\textit{B}}~\cite{astnn} &  & 93.8 & 99.8 & 88.3 & \textbf{70.8} & \textbf{77.3} & 70.5 & \textbf{90.7} & 48.7 & 65.2 & 49.4 & \textbf{96.2} & 51.0  & 42.7 & 51.4 & 36.5 \\
        \rowcolor{gray!20}
        ASTNN\textsubscript{\textit{CL}}  &  &  &  &  & 70.6 & 73.6 & \textbf{72.8} & 79.7 & \textbf{62.3} & \textbf{70.7} & \textbf{57.8} & 91.2 & \textbf{58.2} & \textbf{68.3} & \textbf{55.0} & \textbf{90.1} \\
        \midrule
        CodeGrid\textsubscript{\textit{B}}~\cite{codegrid} &  & 97.8 & 99.6 & 96.1 & 56.0 & 66.7 & 54.1 & \textbf{88.0} & 51.0 & 63.8 & 50.6 & 86.3 & 52.1 & 49.8 & 52.3 & 35.5 \\
        \rowcolor{gray!20}
        CodeGrid\textsubscript{\textit{CL}} &  &  &  &  & \textbf{70.3} & \textbf{75.6} & \textbf{70.2} & 87.0 & \textbf{51.3} & \textbf{64.4} & \textbf{50.8} & \textbf{88.0} & \textbf{64.0} & \textbf{67.2} & \textbf{61.7} & \textbf{73.7}\\
        \bottomrule
    \end{tabular}
    }
\end{table}

\subsection{RQ1 --- Cross-functionality Performance}\label{ssec:rq1_results}
We measure the cross-functionality performance of the selected models and compare it with their performance on the task of detecting clones of seen functionalities. 

\smallskip
\noindent\textbf{Task-Specific Models.}
The evaluation methods \textit{train-bcb-test-scb} and \textit{train-scb-test-bcb} consist of one experiment each (see \Cref{fig:evaluation_methods}) and the performance of the baseline models CodeBERT\textsubscript{\textit{B}}, ASTNN\textsubscript{\textit{B}} and CodeGrid\textsubscript{\textit{B}} is shown in \Cref{table:results}. 
The comparison of the ``\textit{train-bcb-test-scb}'' and ``\textit{train-scb-test-bcb}'' columns of \Cref{table:results} with the ``\textit{on seen functionalities}'' column reveals that the performance of all three models drops up to \num{48}\% in \Fone{} (average \num{31}\%) when the functionalities of the test set do not appear in the training set.

The \textit{one-vs-rest} evaluation method consists of $23$ experiments (\Cref{fig:evaluation_methods}b); hence, we have $23$ tuples of (accuracy, \Fone{}, precision, recall) for each model. 
In \Cref{table:results}, we report the average of the $23$ experiments and provide the full distribution in our replication package~\cite{ours}.
By employing the \textit{one-sample Wilcoxon signed-rank test}~\cite{wilcoxon}, we find that the \Fone{} in the $23$ experiments is lower than the \Fone{} reported in the original papers ($p<0.01$). 
The same holds for accuracy,
while for precision and recall there are cases where the models have low predictive power, leading to nearly \num{100}\% precision with \num{50}\% recall or the opposite.
Since the trade-off between precision and recall can be controlled through the hyperparameter $\tau$ of \Cref{eq:prediction}, we refrain from studying precision and recall independently and study the \Fone{} that combines the two~\cite{stats_book}.
The best-performing task-specific model is CodeBERT, with an average \Fone{} of 71.5\% across the three evaluation methods.

\begin{table}
    \centering
    \scriptsize
    \caption{LLM performance on seen and unseen functionalities of the BCB dataset. LLMs perform worse on unseen functionalities, a drop that  contrastive in-context learning (CL) partially mitigates. $p<.05$ in F1 between baseline and CL is indicated with $^*$.}
    \begin{tabular}{l@{\hskip 0.07in}c@{\hskip 0.07in}c@{\hskip 0.07in}c@{\hskip 0.07in}c@{\hskip 0.07in}l@{\hskip 0.1in}c@{\hskip 0.1in}c@{\hskip 0.1in}c@{\hskip 0.07in}c@{\hskip 0.07in}}
        \toprule
        Evaluation $\rightarrow$ & \multicolumn{4}{c}{on seen functionalities} & \multicolumn{4}{c}{on unseen functionalities} \\
        \cmidrule(lr){2-5} \cmidrule(lr){6-9} 
        Model $\downarrow$ & F1 & A & P & R & F1 & A & P & R  \\
        \midrule
        llama-3.3-70b-versatile\textsubscript{B} & 0.63 & 0.66 & 0.7  & 0.57 & 0.61 & 0.68 & \textbf{0.78} & 0.50  \\
        \rowcolor{gray!20}
        llama-3.3-70b-versatile\textsubscript{CL} &  &  &  &  & \textbf{0.66$^*$} & \textbf{0.69} & 0.73 & \textbf{0.60} \\
        \midrule
        gpt-4o-2024-08-06\textsubscript{B} & 0.64 & 0.68 & 0.73 & 0.57 & 0.59 & 0.63 & 0.70  & 0.51 \\
        \rowcolor{gray!20}
        gpt-4o-2024-08-06\textsubscript{CL} &  &  &  &  & \textbf{0.62$^*$} & \textbf{0.64} & \textbf{0.66} & \textbf{0.59} \\
        \midrule
        deepseek-r1-distill-qwen\textsubscript{B} & 0.70 & 0.70  & 0.69 & 0.70  & 0.68   & \textbf{0.68} & \textbf{0.68} & 0.68 \\
        \rowcolor{gray!20}
        deepseek-r1-distill-qwen\textsubscript{CL} &  &  &  &  & \textbf{0.69} & 0.67 & 0.65 & \textbf{0.73} \\
        \bottomrule
    \end{tabular}
    \label{tab:llm_results}

\end{table}

\smallskip
\noindent\textbf{Generative LLMs.}
We present the LLM results in \Cref{tab:llm_results}, where the models with suffix \texttt{B} indicate the baseline prompt.
The best-performing LLM on unseen functionalities is DeepSeek with 69\% \Fone{}, on par with task-specific models.
All models perform worse on unseen functionalities compared to seen ones, with an average drop of 3\% and a maximum drop of 5\%; 
the reasoning model DeepSeek is affected the least.
The performance drop is smaller than the task-specific models (average 31\%).
This is to be expected: LLMs have orders of magnitude more parameters and are pretrained in a corpus that probably contains many of the functionalities in the dataset, although as unstructured text rather than clone pairs.
Yet, the F1 still drops up to 5\% when the in-context learning examples come from a different functionality.

\begin{tcolorbox}[colback=gray!8!white, colframe=gray!60!black, boxrule=0.5pt, arc=2pt, left=6pt, right=6pt, top=4pt, bottom=4pt]
\begin{minipage}{\textwidth}
\textbf{Finding 1}. Task-specific models achieve an \Fone{} of up to 71.5\% and LLMs up to 69\% on unseen functionalities.
LLMs generalize better to unseen functionalities, experiencing an average \Fone{} drop of 3\%, compared to task-specific models that experience 31\% avg. \Fone{} drop.
\end{minipage}
\end{tcolorbox}

\subsection{RQ2 --- Contrastive Learning Improvements}\label{ssec:rq2_results}

We investigate the use of contrastive learning to improve the performance of existing models on unseen functionalities.
\Cref{table:results} shows the results for the \textit{train-bcb-test-scb} and \textit{train-scb-test-bcb} evaluation methods.
We compare the baseline variation of each model with its respective CL variation, where the best-performing variation is shown in bold. 
Results indicate that the CL variation outperforms the baseline in \Fone{} and accuracy in all six experiments. 
The largest improvement is for ASTNN under the \textit{train-scb-test-bcb} method where contrastive learning increased the F1 by 26\%.
The two cases where a baseline has higher precision is because of low predictive power: the \Fone{} of the CL variation is higher in both cases. The same holds for one case where a baseline has higher recall.

The average of each measure for the \textit{one-vs-rest} method is shown in \Cref{table:results}, while the full distribution is available in our replication package~\cite{ours}.
To compare the CL variation with its baseline counterpart, we employ the \textit{Wilcoxon signed-rank test}~\cite{wilcoxon}.
We find that CL outperforms the baseline in the case of CodeGrid ($p<0.01$) both in accuracy and \Fone{} while for CodeBERT and ASTNN there is no statistically significant difference.
Overall, we find that the CL variation performs better than the baseline in seven out of nine experiments, whereas in two experiments performance is unchanged.

Finally, \Cref{tab:llm_results} shows the results of our generative LLM experiments, where the \texttt{CL} subscript indicates the contrastive prompt. 
To test if the contrastive prompt performs better than the baseline one, we run the Wilcoxon signed-rank test over 10-fold cross-validation \Fone{} scores~\cite{kfold_test}.
We find statistically significant improvement for gpt-4o ($p = .02$, $r=.62$) and llama ($p=.04$, $r=.55$), while DeepSeek experiences no significant improvement, which we attribute to its advanced reasoning.
The average increase in \Fone{} is 3\%, smaller than the task-specific models (9\%).

\begin{tcolorbox}[colback=gray!8!white, colframe=gray!60!black, boxrule=0.5pt, arc=2pt, left=6pt, right=6pt, top=4pt, bottom=4pt]
\begin{minipage}{\textwidth}
\textbf{Finding 2}. Contrastive learning increases the \Fone{} of task-specific models on clones of unseen functionality by up to 26\% (avg. 9\%). Contrastive in-context learning increases the \Fone{} of llama (5\%) and gpt-4o (3\%), with DeepSeek being unaffected.
\end{minipage}
\end{tcolorbox}

We manually investigate the only case where CL performs worse than the baseline in terms of average \Fone{}, although the difference is still not statistically significant and could be just an effect of random noise. This happens to the ASTNN model under the \textit{one-vs-rest} evaluation.
We select three functionalities uniformly at random, namely \texttt{Copy File}, \texttt{GCD}, and \texttt{Test Palindrome}, and analyze the three experiments where the test set contains only these functionalities.
We examine the pairs in the test set which ASTNN\textsubscript{\textit{B}} correctly classified and ASTNN\textsubscript{\textit{CL}} incorrectly classified.
We find $40$ ($20$\%) such pairs for \texttt{Copy File}, of which $37$ are predicted to be non-clones by ASTNN\textsubscript{\textit{CL}} while their ground truth is clones (false-negatives).

By manually inspecting the $37$ false-negatives, we find that $33$ of them are non-clones under 
the Krinke et al.~\cite{bcb_harmful} strict definition of clones which requires clones to implement some functionality as their main or only purpose.  
We discuss this definition at greater length in \Cref{ssec:robustness_testing}.
For example, in a clone (according to BCB ground truth) pair, one snippet copies a file from one location to another while the second snippet additionally imports a certificate into a Java KeyStore, which involves a series of file operations.
ASTNN\textsubscript{\textit{CL}} classified this pair as a non-clone and a similar pattern is followed in $33$ out of the $37$ false-negatives.
By inspecting the pairs for the \texttt{GCD} and \texttt{Test Palindrome} functionalities, \textit{we find that all the pairs labeled as clones in the dataset are actually clones under the strict definition}.
Moreover, ASTNN\textsubscript{\textit{CL}} outperforms ASTNN\textsubscript{\textit{B}} when the test set contains these two functionalities.
This provides evidence for why CL did not improve ASTNN performance in the \textit{one-vs-rest} evaluation:  it learned a stricter clone definition while the test set uses a more permissive definition for some functionalities, like \texttt{Copy File}.
In \Cref{sec:discussion}, we propose areas for future work based on these qualitative results.

Another result comes from the interpretation of the \textit{one-vs-rest} experiments, which measure
how well models can detect clones of functionality $f$ given that they have never seen clones of $f$  in training. 
For example, CodeBERT\textsubscript{\textit{CL}} achieves $95\%$ accuracy when tested on the functionality \texttt{Shuffle Array in Place}, while ASTNN\textsubscript{\textit{CL}} and CodeGrid\textsubscript{\textit{CL}} achieve $73\%$ and $52\%$ respectively. 
Similarly, ASTNN\textsubscript{\textit{CL}} is the only model that achieves near-perfect performance when tested on the \texttt{Open URL in System Browser} functionality with an accuracy of $96\%$ and CodeGrid\textsubscript{\textit{CL}} is the only model that performs strongly on the \texttt{Secure Hash} functionality with an accuracy of $94\%$. 
This observation suggests that the inherently complex nature of semantic clones may not be fully captured with a single code representation.

\begin{table}
    \centering
    \scriptsize
    \caption{Performance of the three task-specific models on the OJ dataset. The contrastive learning variation outperforms the baseline variation in all three models in \Fone{}.}\label{table:results_robustness_testing}
    \begin{tabular}{@{}lcccc|cccc}
        \toprule
         Evaluation $\rightarrow$ & \multicolumn{4}{c}{train-oj-test-scb} & \multicolumn{4}{c}{train-scb-test-oj} \\
        \cmidrule(lr){2-5} \cmidrule(lr){6-9} 
        Model $\downarrow$ & A & \Fone{} & P & R & A & \Fone{} & P & R \\
        \midrule
        CodeBERT\textsubscript{\textit{B}}~\cite{codebert} & 56.6 & 27.9 & 16.8 & 82.0 & \textbf{87.7} & 17.9 & 20.5 & \textbf{16.0} \\
        \rowcolor{gray!20}
        CodeBERT\textsubscript{\textit{CL}}  & \textbf{79.0} & \textbf{76.1} & \textbf{66.6} & \textbf{88.7} & 69.2 & \textbf{23.3} & \textbf{71.0} & 13.9 \\
        \midrule
        ASTNN\textsubscript{\textit{B}}~\cite{astnn} &  66.4 & 59.5 & \textbf{74.9} & 49.3 & 75.6 & 23.1 & 14.5 & \textbf{55.7} \\
        \rowcolor{gray!20}
        ASTNN\textsubscript{\textit{CL}} & \textbf{72.1} & \textbf{70.5} & 74.7 & \textbf{66.7} & \textbf{91.8} & \textbf{37.3} & \textbf{37.5} & 37.2 \\
        \midrule
        CodeGrid\textsubscript{\textit{B}}~\cite{codegrid} & 50.0 & \textbf{66.7} & 50.0 & \textbf{100} & 57.5 & 14.5 & 8.4 & \textbf{55.0} \\
        \rowcolor{gray!20}
        CodeGrid\textsubscript{\textit{CL}}  & \textbf{54.0} & 57.9 & \textbf{53.4} & 63.1 & \textbf{68.7} & \textbf{14.9} & \textbf{9.0} & 42.8\\
        \bottomrule
    \end{tabular}
\end{table}

\subsection{Evaluating on Strict Clones}\label{ssec:robustness_testing}
To strengthen our findings, we want to generalize our results to datasets other than BCB, which uses a partial definition of clones. 
\Cref{def:clones} leaves what constitutes the \textit{same functionality} intentionally imprecise.  
BCB considers two snippets to implement the same functionality (and therefore to be clones) when they both implement one of the 43 predefined functionalities, regardless of any additional functionalities those snippets may implement.
For example, consider two code snippets: the first one copies a file to a local directory; 
the second one opens an FTP connection, copies a file to the FTP server, and closes the  FTP connection. 
According to BCB, the two snippets form a clone pair. 

Krinke et al.~\cite{bcb_harmful} consider another, much stricter definition of clones: \emph{two snippets form a clone pair if they implement functionality $f$ as their main or only purpose.}
Armed with this definition, they 
manually examined \num{100} clones of the \texttt{Copy File} functionality to check whether they are still clones.
They find that \num{86}\% of the pairs would not be considered clones under this strict definition. 
They only inspected clones of the \texttt{Copy File} functionality, which is usually seen as part of bigger methods, in contrast to other functionalities of BCB like \texttt{Fibonacci} or \texttt{GCD}, which are more likely to be seen as standalone methods as we observed in~\Cref{ssec:rq2_results}.

There is no universally accepted, decidable definition of clone, perhaps as a consequence of treading in undecidable waters.  For instance, Krinke \emph{et al.}'s strict definition leaves open the question of just what constitutes ``main or only''.  Perhaps to avoid difficulties such as these, some think the definition should depend on the end-user application~\cite{baxter1998clone}.
In any event, we want our results to be independent of the definition used by BCB. 
For this reason, we re-run all our experiments after replacing BCB with the OJ dataset, which follows Krinke \emph{et al.}'s strict definition. 

The results for the task-specific models are summarized in \Cref{table:results_robustness_testing}.
We see that, similarly to \Cref{table:results}, CL outperforms the baseline in \Fone{} in 5/6 experiments, while in the other experiment the baseline simply predicts everything to be a clone due to the class imbalance of OJ, leading to higher \Fone{} but lower accuracy.
The \textit{one-vs-rest} evaluation does not apply to the OJ dataset because OJ does not contain non-clone pairs for each functionality.
We also note that the low precision, recall, and \Fone{} are due to the class imbalance of the OJ dataset described in \Cref{sec:background}.
Regarding LLMs, we notice that they perform particularly well on OJ, with \Fone{} ranging from 95\% to 99\% (full table in our material~\cite{ours}), regardless of the prompting strategy.
This indicates that LLMs perform better under a strict definition of clones, as a result of dealing with less ambiguity.
This set of experiments broadens the scope of our experimental conclusions.

\section{Discussion}\label{sec:discussion}
In this section, we discuss the implications of our findings. 
With our results, we hope to \emph{steer the focus toward evaluations that test the models on clones of unseen functionality}. 
Our analysis provides novel evidence that this evaluation is more aligned with real-world scenarios, making it a prominent option for evaluating future models.

Even after the contrastive learning improvements, there is more headroom for improvement in both task-specific models and LLMs. 
For task-specific models, for example, future work can investigate whether and how adding more functionalities to a dataset can increase the cross-functionality performance. 
From our experiments, we see that training the task-specific models on $23$ functionalities of BCB$_{s'}$ is not enough. 
Although BCB can be expanded, the process of doing so is manual, thus 
raising the question of whether generative AI can be employed to create large datasets, as was recently attempted successfully with GPT~\cite{gptclonebench,roy2023unveiling}.
Going a step further, we may need to rethink whether the optimal structure of a semantic clone dataset should be functionality-based like BCB, OJ, and GCJ in the first place. 
The SCB dataset that was recently proposed is not functionality-based and contains \num{1000} Java clone pairs, each one implementing a unique functionality.
However, the small size of only \num{1000} clones hinders its use for training task-specific models.

Our qualitative analysis suggests that CL in task-specific models may not be effective in predicting partial clones.
Further investigation is required to understand how the definition of clones affects the performance of different architectures.
For example, it seems worth investigating whether and how changing the hyperparameters $m$ and $\tau$ of \Cref{eq:contrastive_loss} and \Cref{eq:prediction} respectively can improve the performance of CL on non-strict clones.
Regarding LLMs, we find that while they achieve near-perfect performance on strict clones, they also struggle on partial clones, leaving headroom for improvement through enhanced multi-step prompting or finetuning on partial clone pairs. 

Our application of \emph{CL increases cross-functionality performance of existing models in 9/12 experiments}. 
Although CL has been used for detecting clones before, we are the first to show that it improves the performance of existing models in the unseen functionality setting.
Given that it is a model-agnostic technique that can work with any model, future research in clone detection can incorporate this finding. 
The optimal architecture and hyperparameters for CL can also be the focus of further studies.

Finally, we find that some functionalities are only captured with one code representation. 
Although the sample is small, it suggests that a single representation of code (text, AST, image) may not be sufficient to universally solve the complex problem of semantic clone detection. This reveals a potentially fertile area of future research toward the development of ensemble models that combine information from text, AST, and image sources of code.

\section{Threats to Validity}\label{sec:threatsToValidity}

Here, we discuss the threats to the validity of our study.
\noindent\textbf{Model and Dataset Selection.}
We acknowledge the threat due to the selected models and limit our conclusions to these models only. 
To minimize this threat, we selected models based on a meaningful distinguishing feature: the input format, selecting a text-based, an AST-based, and an image-based model.
We also incorporated LLMs, further broadening our scope.
A threat to validity in LLM evaluation is data leakage, or memorization of some training data~\cite{memorization}.
The fact that LLMs perform slightly worse than task-specific models in BCB reduces the threat of memorization, but memorization is still possible in the OJ dataset. 
To mitigate this threat, we also use task-specific models that we train from scratch to reach our conclusions.

Existing datasets for semantic clone detection are a threat to validity~\cite{wysiwim,bcb_harmful,astnn,krinke2025misuse}.
To minimize the functionality imbalance of BCB, which is a threat to validity in other works~\cite{wysiwim,bcb_harmful}, we introduced a balanced version.
It was also shown that BCB\textsubscript{v2} contains unlabeled clones~\cite{bcb_harmful}, making the precision potentially inaccurate. 
To mitigate this, researchers measure precision by manually assessing a small subset of the dataset~\cite{cclearner}. 
We do not follow this process because our goal is not an exact precision score, but rather empirical evidence of (a) weak cross-functionality performance and (b) improved cross-functionality performance with CL. 
Other measures like accuracy and \Fone{} provide evidence to minimize the threat of a potentially inaccurate calculation of precision.
Finally, the granularity of some BCB functionalities is low, like \texttt{Copy File} whose implementation is only three lines in Java~\cite{bcb_harmful}.
Our experiments on OJ, whose functionalities are not trivial, mitigate this threat.

\smallskip
\noindent\textbf{Additional Sources of Distribution Shift.}
To measure cross-functionality performance of task-specific models, we introduced three evaluation methods. Two of them involve training a model on a dataset and evaluating it on another one. Since the functionalities of the two datasets differ, these methods inherently capture cross-functionality performance. 
However, there may be additional sources of distribution shift between the two datasets, like different coding styles or application domains, that could lead to lower performance.
This poses the threat that the performance drop is not \textit{exclusively} due to weak cross-functionality performance.
We argue that based on how the datasets are constructed, coding styles or application domains should not play a significant role: BCB contains code snippets from \num{25}K projects varying in application domain and authorship. Similarly, SCB contains \num{1}K stackoverflow answers on different topics from different authors. 
This diversity should prevent a model trained on BCB from overfitting to a specific coding style or domain.

Nevertheless, to further mitigate this threat, we have also evaluated cross-functionality performance within the BCB dataset, in our \textit{one-vs-rest} evaluation method, where we keep one functionality of the BCB dataset in the test set and use the rest of the functionalities for training.
The results from all three settings agree; this triangulation increases our confidence that there is weak performance on clones of unseen functionality.

\smallskip
\noindent\textbf{Resampling BCB.}
We introduced BCB$_{s'}$ by resampling BCB\textsubscript{v2} in a functionality-aware manner, to mitigate the functionality imbalance~\cite{wysiwim,bcb_harmful}.
We used the balanced BCB$_{s'}$ in our experiments,
which introduces the threat that the performance drop reported in RQ1 (\Cref{ssec:rq1_results}) may be due to different subsets of the BCB dataset.
To mitigate this threat, we conduct three more experiments:
we train and evaluate the three models using the evaluation method of the original papers on the BCB$_{s'}$ dataset.
Results show that the performance of all three models on BCB$_{s'}$ is equally high as in the original papers (\Fone{} is \num{94.0}\%, \num{97.8}\%, and \num{92.5}\% respectively). 
This gives further confidence that the performance drop we observe in RQ1 (cross-functionality performance compared to the performance reported in the original papers) should not be an effect of the different BCB version but of the weak generalization of the models to clones of unseen functionality.

\section{Conclusion}\label{sec:conclusion}
We present an in-depth analysis of the cross-functionality performance of clone detection models. 
We start by highlighting the relevance of detecting clones of unseen functionality to real-world applications.
This motivates us to study the evaluation methodology of state-of-the-art models, finding that they were evaluated on the task of detecting clones of seen functionalities.
By testing six models on clones of unseen functionality, we show that there is a significant drop in performance.
This leaves ample room for future research toward achieving high cross-functionality performance.

Finally, we propose and evaluate the use of contrastive learning to increase the cross-functionality performance of task-specific models by replacing their final classifier with a contrastive classifier.
In addition, we introduce an analogous technique for clone detection with generative LLMs, called contrastive in-context learning.
We provide evidence that, in most cases, contrastive learning improves cross-functionality performance, making it a prominent option for future research.

\section*{Acknowledgments}
K. Kitsios, F. Sovrano, and A. Bacchelli gratefully acknowledge the support of the Swiss National Science
Foundation through the SNSF Project 200021\_197227.

\begingroup
\footnotesize
\bibliographystyle{IEEEtranN}
\bibliography{references}
\endgroup

\end{document}